\documentclass[final,5p,fleqn]{elsarticle}

\usepackage{amssymb}
\usepackage[mathscr]{eucal}
\usepackage{amsmath}
\usepackage{graphicx}
\usepackage{pstricks}
\usepackage{wrapfig}
\newcommand{\e}{\mathbf}

\begin{document}

\begin{frontmatter}

\title{Spin correlations near the surface \\ of a three-dimensional Heisenberg antiferromagnet}

\author{N. Voropajeva and A. Sherman}

\address{Institute of Physics, University of Tartu, Riia 142, 51014 Tartu, Estonia}

\begin{abstract}
Nearest-neighbor spin correlations are considered near the surface of a semi-infinite spin-$\frac12$ Heisenberg antiferromagnet on a simple cubic lattice. In the spin-wave approximation, the excitation spectrum of this model involves bulk modes -- standing spin waves and a quasi-two-dimensional mode of surface spin waves. These latter excitations eject the bulk excitations from the surface region thus dividing the antiferromagnet into two parts with different excitations. As a result absolute values of the spin correlations near the surface exceed the bulk value. In the surface region, the pattern of spin correlations resembles the comb structure recently obtained for the two-dimensional case.
\end{abstract}

\begin{keyword}
Semi-infinite Heisenberg antiferromagnet \sep magnetic excitations \sep spin correlations 

\PACS 75.10.Jm \sep 75.30.Ds
\end{keyword}

\end{frontmatter}

\section{Introduction}
The influence of boundaries on the spectrum and observables of the quantum Heisenberg antiferromagnet has been studied in one \cite{mahajan,takigawa} and two \cite{hoglund,metlitski,pardini} dimensions. One of the results obtained in the two-dimensional (2D) case with the use of Monte Carlo simulations \cite{hoglund}, the spin-wave approximation \cite{metlitski} and the series expansion  \cite{pardini} is the increased absolute values of the nearest-neighbor spin correlations near the boundary of the antiferromagnet. With distance from the surface the correlations tend rapidly to their bulk value revealing some oscillations. The arising pattern of spin correlations was called the comb structure. It was argued that the increased surface correlations can be a manifestation of a short-range valence-bond-solid ordering in the N\'eel phase \cite{metlitski,pardini}.

In this paper, we study the influence of the surface on the nearest-neighbor spin correlations in the three-di\-mensional (3D) Heisenberg antiferromagnet on a simple cubic lattice. Using the spin-wave approximation we found the distribution of the spin correlations which resembles the comb-like structure of the 2D case: the correlations on the surface and between the surface and the second to the surface layers are in absolute value larger than in the bulk, while in the second layer and between the second and the third layers the correlations are smaller than in the bulk. In the 3D case, the differences between the surface and the bulk values of the correlations are smaller than in the 2D case and decay more rapidly with distance from the surface -- starting from the third layer the correlations differ only slightly from their bulk value.

We relate the appearance of such a spin correlation pattern to the peculiar spectrum of spin excitations near the surface of the antiferromagnet. The spectrum involves bulk modes -- standing spin waves and a quasi-two-dimen\-sional mode of surface spin waves. These latter excitations are observed in the two surface layers, and they eject the bulk excitations from this region \cite{sherman}. Thus the antiferromagnet appears to be divided into two regions with different spin excitations. As known, in the case of an infinite 2D Heisenberg antiferromagnet the nearest-neighbor spin correlations are larger in absolute value than in the same antiferromagnet in the 3D case. Hence the quasi-two-dimensional surface mode yields stronger correlations than the 3D bulk modes, which explains larger values of the correlations near the surface and the comb structure.

It is worth noting that the description of perturbations introduced by the surface into the magnon spectrum is in many respects similar to the problem of a local defect in a crystal \cite{lifshits}. In this latter problem the crystal is also divided into two regions with different elementary excitations -- a vicinity of the defect with localized states and the rest of the crystal with bulk states.

\section{Model and its elementary excitations}
The axes are chosen in such a way that the antiferromagnet is in the half-space $l_x\geqslant0$. Here sites of a 3D simple cubic lattice are labeled by the three coordinates $l_x,l_y,l_z$ and the lattice spacing is set as the unit of length. The system is described by the Hamiltonian
\begin{equation}\label{1}
    H=\frac{J}{2}\sum_{\e{l}\e{a}}\sum_{l_x\geqslant0}
    \e{S}_{\e{l+a},l_x}\e{S}_{\e{l}l_x}+
    J\sum_{\e{l},l_x\geqslant0}\e{S}_{\e{l},l_x+1}\e{S}_{\e{l}l_x}\,,
\end{equation}
where $\e l=(l_y,l_z)$, $\e a=(\pm1,0),(0,\pm1)$ are four unit vectors which connect nearest neighbor sites in the $YZ$ plane, and $\e S_{\e L}$ is the $S=1/2$ spin operator.

Since for low temperatures the system has the long-range antiferromagnetic order, its low-lying elementary excitations can be described in the spin wave approximation,
\begin{equation}\label{2}
    S^z_{\e{L}}=e^{i\e{\Pi
    L}}\left(\frac12-b^\dag_\e{L}b_{\e{L}}\right), \quad
    S^\pm_\e{L}=P^{\pm}_\e{L}b_\e{L}+P_\e{L}^{\mp}
    b^\dag_\e{L},
\end{equation}
where the spin-wave operators $b_{\e L}$ and $b^\dag_{\e L}$ satisfy the Boson commutation relations and
\begin{equation*}
    \e{\Pi}=(\pi,\pi,\pi),
    \quad P^\pm_\e{L}=\frac12\left(1\pm e^{i\e{\Pi L}}\right), \quad
    \e L=(l_x,l_y,l_z).
\end{equation*}
If we substitute Eq.~\eqref{2} into Eq.~\eqref{1}, drop constant terms and terms containing more than two spin-wave operators, we obtain
\begin{eqnarray}\label{3}\nonumber
    H&=& 3J\sum_{\e k l_x\geqslant0}\left(1-\frac16\delta_{l_x 0}\right)b^\dag_{\e k l_x}b_{\e k l_x}\\\nonumber
    &+&J\sum_{\e k l_x\geqslant0}\gamma^{(2)}_{\e k}\left(b_{\e{k}l_x}b_{\e{-k},l_x}+b^\dag_{\e k l_x}b^\dag_{\e{-k},l_x}\right)\\
    &+&\frac J2\sum_{\e k l_x\geqslant0}\left(b_{\e k l_x}b_{\e{-k},l_x+1}+b^\dag_{\e k l_x}b^\dag_{\e{-k},l_x+1}\right).
\end{eqnarray}
Here we took into account the translational invariance of
Hamiltonian \eqref{1} in the $YZ$ plane and used the Fourier transformation
\begin{equation*}\label{}
    b_{\e k l_x}=\frac{1}{\sqrt N}\sum_{\e l}e^{i\e {kl}}b_{\e l l_x},
\end{equation*}
where $\e k$ is a 2D wave vector and $N$ is the number of sites in the periodic $YZ$ plane. In Eq.~\eqref{3},
$\gamma^{(2)}_{\e k}=\frac12[\cos(k_y)+\cos(k_z)]$.

To diagonalize quadratic form \eqref{3}, we
use the Bogo\-liu\-bov-Tyablikov transformation
\cite{tyablikov} which in the pre\-sent case reads
\begin{equation}\label{4}
    b_{\e k l_x}=\sum_{k_x} \left(u_{\e k l_x k_x}\beta_{\e k k_x}+v_{\e k l_x k_x}\beta^\dag_{-\e k,k_x}\right),
\end{equation}
with the inverse transformation
\begin{equation}\label{5}
    \beta_{\e k k_x}=\sum_{l_x\geqslant0}(u^{*}_{\e k l_x k_x} b_{\e k l_x}-v_{-\e k,l_x k_x}b^\dag_{-\e k, l_x}).
\end{equation}
Since the operators $\beta_{\e k k_x}$ and $\beta^\dag_{\e k k_x}$ satisfy the Boson commutation relations, the following conditions are imposed on the coefficients $u_{\e k l_x k_x}$ and $v_{\e k l_x k_x}$:
\begin{equation}
\begin{array}{lcl}\label{6}
    \sum\limits_{k_x}\left(u_{\e k l_x k_x}u^{*}_{\e k l'_x k_x}-v_{\e k l_x k_x}v^{*}_{\e k l'_x k_x}\right)&=&\delta_{l_x l'_x},\\
    \sum\limits_{k_x}\Bigl(u_{\e k l_x k_x}v_{-\e k,l'_x k_x}-v_{\e k l_x k_x}u_{-\e k,l'_x k_x}\Bigr)&=&0,\\
    \sum\limits_{l_x\geqslant0}\left(u_{\e k l_x k_x}u^{*}_{\e k l_x k_x'}-v_{-\e k,l_x k_x'}v^{*}_{-\e k,l_x k_x}\right)&=&\delta_{k_x k_x'},\\
    \sum\limits_{l_x\geqslant0}\left(u^{*}_{-\e k,l_x k_x}v_{-\e k l_x k_x'}-v_{\e k l_x k_x}u^{*}_{\e k l_x k_x'}\right)&=&0.
\end{array}
\end{equation}
In the new representation the Hamiltonian \eqref{3} has the diagonal form,
\begin{equation*}
    H=\sum_{\e k k_x}E_{\e k k_x}\beta^\dag_{\e k k_x}\beta_{\e k k_x}+\mathrm{const}.
\end{equation*}
If we use Eqs.~\eqref{3}-\eqref{6} in the relation
$$[\beta_{\e k k_x},H]=E_{\e k k_x}\beta_{\e k k_x}$$ we find the system of equations for the determination of the coefficients $u_{\e k l_x k_x}$, $v_{\e k l_x k_x}$ and the energy $E_{\e k k_x}$,
\begin{eqnarray}
    &&E_{\e k k_x}u^{*}_{\e k l_x k_x}=
    3J\left(1-\frac16 \delta_{l_x 0}\right)u^{*}_{\e k l_x k_x}\nonumber\\
    &&\quad+2J\gamma^{(2)}_{\e k}v_{-\e k,l_x k_x}+\frac J2\left(v_{-\e k,l_x+1,k_x}+v_{-\e k,l_x-1,k_x}\right), \nonumber\\
    &&\label{7}\\[-1ex]
    &&-E_{\e k k_x}v_{-\e k,l_x k_x}=
    3J\left(1-\frac16\delta_{l_x 0}\right)v_{-\e k,l_x k_x}\nonumber\\
    &&\quad+2J\gamma^{(2)}_{\e k} u^{*}_{\e k l_x k_x}+\frac J2\left(u^{*}_{\e k,l_x+1,k_x}+u^{*}_{\e
    k,l_x-1,k_x}\right),\nonumber
\end{eqnarray}
with the boundary conditions
\begin{equation}\label{8}
    u^{*}_{\e k,l_x=-1,k_x}=0, \quad v_{-\e k,l_x=-1,k_x}=0.
\end{equation}

If, for the time being, terms proportional to $\delta_{l_x0}$ in Eq.~\eqref{7} are neglected the solutions for this simplified system of equations can be written in the form
\begin{equation*}\label{}
    u^{*}_{\e k l_x k_x} \sim e^{\varkappa(k_x) l_x},\quad
    v_{\e -k,l_x k_x}\sim e^{\varkappa(k_x)l_x},
\end{equation*}
where $\varkappa$ has to be either purely imaginary or real for the
energy $E_{\e k k_x}$ be real. Note that a linear combination of such solutions corresponding to the same eigenenergy [for a fixed pair $(\e k,-\e k)$] is also a solution of system \eqref{7}. Solutions with real $\varkappa$-s do not satisfy boundary conditions \eqref{8}. Only a linear combination of two solutions with imaginary and opposite in sign $\varkappa$-s, $\varkappa=\pm i k_x$, a standing wave, satisfies  system \eqref{7} and conditions \eqref{8},
\begin{eqnarray}
    u^{*}_{\e k l_x k_x}&=&A_{\e k k_x}\sin[k_x(l_x+1)],\nonumber\\
    v_{-\e k,l_x k_x}&=&B_{\e k k_x}\sin[k_x(l_x+1)],\nonumber\\
    A_{\e k k_x} &=& \sqrt{\frac{2}{\pi}}\frac{3J+E_{\e k k_x}}{\sqrt{\left(3J+E_{\e k k_x}\right)^2-\left(3J\gamma^{(3)}_{\e k k_x}\right)^2}},\label{9}\\
    B_{\e k k_x} &=& -\sqrt{\frac{2}{\pi}}\frac{3J\gamma^{(3)}_{\e k k_x}}{\sqrt{\left(3J+E_{\e k k_x}\right)^2-\left(3\gamma^{(3)}_{\e k k_x}\right)^2}},\nonumber\\
    E_{\e k k_x}&=&3J\sqrt{1-\left(\gamma^{(3)}_{\e k k_x}\right)^2}, \nonumber
\end{eqnarray}
where $\gamma^{(3)}_{\e k k_x}=\frac13[\cos(k_x)+\cos(k_y)+\cos(k_z)]$ and $k_x$ varies continuously in the range $(0,\pi)$. Thus, in the above equations sums over $k_x$ and Kronecker symbols containing $k_x$ have to be substituted with integrals and the Dirac delta functions, respectively.

If in Eq.~\eqref{7} we take into account the previously dro\-pp\-ed terms proportional to $\delta _{l_x 0} $, the simple exponential solutions are inapplicable. To solve this more intricate problem, we modify the method used for the local defect problem \cite{lifshits}. Let us introduce the two-component operator
\begin{equation*}\label{}
    \hat B_{\e k l_x}=
    \left(
      \begin{array}{c}
        b_{\e k l_x} \\
        b^\dag_{-\e k l_x} \\
      \end{array}
    \right)
\end{equation*}
and define the matrix retarded Green's function
\begin{equation}\label{10}
    \hat D(\e k t l_x l'_x)=-i\theta(t)\left\langle\left[\hat B_{\e k l_x}(t),\hat B^\dag_{\e k l'_x}\right]\right\rangle,
\end{equation}
where $\hat B_{\e k l_x}(t)=e^{i H t}\hat B_{\e k l_x}e^{-H t}$
with $H$ determined by Eq. \eqref{3}. If we define $\hat
D^{(0)}(\e k t l_x l'_x)$ as Green's function corresponding to Hamiltonian \eqref{3} without the term proportional to $\delta_{l_x
0}$, $\hat D(\e k t l_x l'_x)$ can be
expressed as
\begin{equation*}
    \hat D(\e k t l_x l'_x)=\hat D^{(0)}(\e k t l_x l'_x)
\end{equation*}
\begin{equation}\label{11}
    \hspace{2cm}-\frac J2\int\limits_{-\infty}^\infty dt' \hat D^{(0)}(\e k,t-t',l_x 0)\hat D(\e k t' 0 l'_x).
\end{equation}
After the Fourier transformation we obtain the solution,
\begin{equation*}
    \hat D(\e k \omega l_x l'_x)=\hat D^{(0)}(\e k \omega l_x l'_x)
    -\frac J2 \hat D^{(0)}(\e k \omega l_x 0)
\end{equation*}
\begin{equation}\label{12}
    \hspace{2cm}\times\left[\hat\tau_0+\frac J2 \hat D^{(0)}(\e k \omega 0 0)\right]^{-1}\hat D^{(0)}(\e k \omega 0 l'_x),
\end{equation}
where $\hat\tau_0$ is a $2\times2$ unit matrix. Except for the matrix form and the dependence on \textbf{k}, Eq.~(12) is similar in form to the equation for Green's function of a crystal with a local defect \cite{lifshits}.

Green's function $\hat D^{(0)}(\e k \omega l_x l'_x)$ is easily calculated with the use of Eq.~\eqref{9},
\begin{eqnarray}
    &&\hat D^{(0)}(\e k \omega l_x l'_x)=
    \int_0^\pi\!\! dk_x \sin[k_x(l_x+1)\sin[k_x(l'_x+1)]\nonumber\\
    &&\quad\times\left(\frac{1}{\omega-E_{\e k k_x}+i\eta}\hat P_{\e k k_x}-\frac{1}{\omega+E_{\e k k_x}+i\eta}\hat Q_{\e k k_x}\right), \nonumber\\
    &&\label{13}\\[-1ex]
    &&\hat P_{\e k k_x}=\left(
                   \begin{array}{cc}
                     A^2_{\e k k_x} & A_{\e k k_x}B_{\e k k_x} \\
                     A_{\e k k_x}B_{\e k k_x} & B^2_{\e k k_x} \\
                   \end{array}
                 \right),\nonumber\\
    &&\hat Q_{\e k k_x}=\left(
                   \begin{array}{cc}
                     B^2_{\e k k_x} & A_{\e k k_x}B_{\e k k_x} \\
                     A_{\e k k_x}B_{\e k k_x} & A^2_{\e k k_x} \\
                   \end{array}
                 \right),\nonumber
\end{eqnarray}
where $\eta=+0$.
%

\begin{figure}[t]
\begin{center}
\includegraphics[width=0.8\linewidth]
{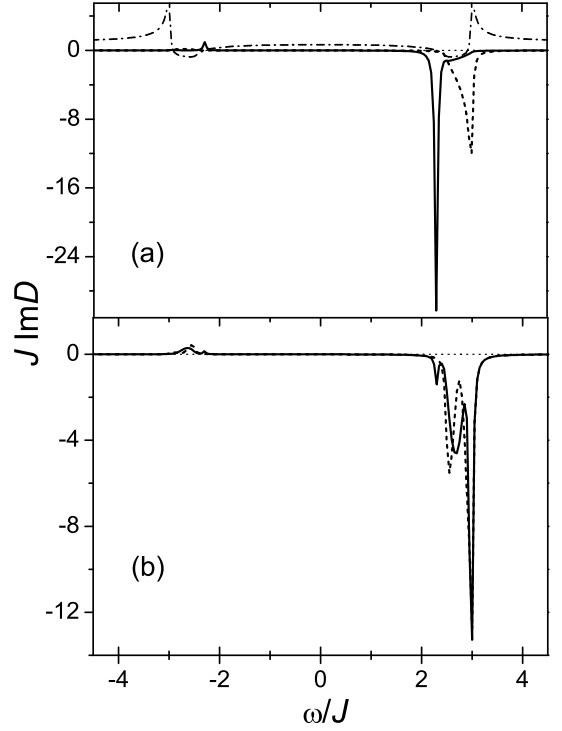} \caption{The imaginary parts of Green's functions $D_{11}(\e k\omega l_x l_x)$ (the solid lines) and $D_{11}^{(0)} (\e k\omega l_x l_x)$ (the dashed lines) for $l_x =0$ (a) and $l_x =2$ (b). $\e k=\left(0,0.6\pi\right)$. In part (a), the dash-dotted line demonstrates the real part of the denominator in the second term on the right-hand side of Eq.~(12).} \label{fig1}
\end{center}
\end{figure}

The poles of Green's function $\hat {D}^{(0)}({\rm {\bf k}}\omega l_x {l}'_x)$ correspond to standing spin waves \eqref{9}. Apart from them  Green's function $\hat {D}({\rm {\bf k}}\omega l_x {l}'_x )$ may have poles connected with the second term on the right-hand side of Eq.~\eqref{12}. The imaginary parts of Green's functions $\hat {D}^{(0)}({\rm {\bf k}}\omega l_x l_x )$ and $\hat {D}({\rm {\bf k}}\omega l_x l_x )$ are shown in Fig.~\ref{fig1} for different distances from the surface. On the surface, $l_x =0$, the spectrum
$\mbox{Im }\hat {D}({\rm {\bf k}}\omega l_x l_x )$ is dominated by the peak arising from the second term on the right-hand side of Eq.~\eqref{12}. Indeed, as seen from Fig.~1(a), the peak frequency coincides with a zero of the denominator in this term. Notice that the standing spin waves manifest themselves as a weak shoulder on the high-frequency side of the peak -- the excitation connected with it ejects the bulk modes, the standing waves, from the surface layer. The similar situation is observed in the second layer. However, as seen from Fig.~1(b), already in the third layer the peak is weak and the spectrum is dominated by the continuum of standing waves. Thus, the
antiferromagnet is divided into two regions with different spin excitations. The two layers near the surface is the location of the mode connected with the pole of the second term in Eq.~\eqref{12}. This mode has the dispersion of the 2D spin waves,
\begin{equation*}\label{}
    \omega _{\rm {\bf k}}
    =2{J}'\sqrt {1-\left( {\gamma _{\rm {\bf k}}^{(2)} }\right)^2},
\end{equation*}
with a somewhat increased exchange constant ${J}'\approx 1.23J.$ This mode is termed the surface spin-wave mode. Excitations of the rest of the crystal are standing spin waves with dispersion \eqref{9}. Notice that this picture is in many respects similar to the situation in the problem of a local defect \cite{lifshits}: if local states arise near the defect, they eject bulk states from the
defect region.

\section{Spin correlations}
Nearest-neighbor spin correlations can be expressed in terms of the spin-wave operators with the use of Eq.~\eqref{2} and the translational invariance of Hamiltonian \eqref{1}
\begin{eqnarray}\label{14}\nonumber
    \langle \e S_{\e L}\e S_{\e L'}\rangle=
    \frac{1}{2N}\sum_{\e k}\biggl\{2\cos[\e k(\e l-\e l')]\langle b_{\e k l_x}b_{-\e k,l'_x}\rangle\\
    +\langle b^\dag_{\e k l_x}b_{\e k l_x}\rangle+\langle b^\dag_{\e k l'_x}b_{\e k l'_x}\rangle\biggr\}-\frac14.
\end{eqnarray}
The correlation functions in Eq.~\eqref{14} are connected with Green's function \eqref{10} by the relation
\begin{eqnarray}\label{15}\nonumber
    \left\langle \hat B_{\e k l_x}\hat B^\dag_{\e k l'_x}\right\rangle=
    \int\limits_{-\infty}^{\infty} \frac{d\omega}{2\pi}\frac{\mathrm{Im}[\hat D(\e k \omega l_x l'_x)+\hat D^T(\e k \omega l'_x l_x)]}{e^{-\beta \omega}-1}\\
    -i\int\limits_{-\infty}^{\infty} \frac{d\omega}{2\pi}\frac{\mathrm{Re}[\hat D(\e k \omega l_x l'_x)-\hat D^T(\e k \omega l'_x l_x)]}{e^{-\beta \omega}-1},
\end{eqnarray}
where $D^T_{ij}(\e k \omega l_x l'_x)=D_{ji}(\e k \omega l_x l'_x)$. Bearing in mind the property of Green's function \eqref{12}
\begin{equation*}\label{}
    D_{ij}(\e k \omega l_x l'_x)=D_{ji}(\e k \omega l'_x l_x),
\end{equation*}
relation \eqref{15} is considerably simplified,
\begin{equation}\label{16}
    \left\langle \hat B_{\e k l_x}\hat B^\dag_{\e k l'_x}\right\rangle=
    \int_{-\infty}^\infty \frac{d\omega}{\pi}\frac{\mathrm{Im}[\hat D(\e k \omega l_x l'_x)]}{e^{-\omega\beta}-1}.
\end{equation}

Let us consider the nearest-neighbor spin correlations parallel and perpendicular to the surface,
\begin{equation}\label{17}
    C_L=\left\langle \e S_{\e l+\e a,l_x}\e S_{\e l l_x}\right\rangle,\quad
    C_T=\left\langle \e S_{\e l,l_x+1}\e S_{\e l l_x}\right\rangle.
\end{equation}
For $T=0$ with the use of Eq.~\eqref{16} these spin correlations read
\begin{eqnarray}
    C_L&=&-\frac{1}{N}\sum_{\e k} \cos(k_y)\int\limits_{0}^{\infty}\frac{d\omega}{\pi}\mathrm{Im} D_{12}(\e k \omega l_x l_x)\nonumber\\
    &-&\frac{1}{N}\sum_{\e k}\int\limits_{0}^{\infty}\frac{d\omega}{\pi}\mathrm{Im}D_{22}(\e k \omega l_x l_x)-\frac14,\nonumber\\
    C_T&=&-\frac{1}{N}\sum_{\e k}\int\limits_0^{\infty}\frac{d\omega}{\pi}\mathrm{Im}D_{12}(\e k \omega l_x,l_x+1)\label{18}\\
    &-&\frac{1}{2N}\sum_{\e k}\int\limits_0^{\infty}\frac{d\omega}{\pi} \mathrm{Im}D_{22}(\e k \omega,l_x+1,l_x+1)\nonumber\\
    &-&\frac{1}{2N}\sum_{\e k}\int\limits_0^{\infty}\frac{d\omega}{\pi}\mathrm{Im} D_{22}(\e k \omega l_x l_x)-\frac14.\nonumber
\end{eqnarray}
\begin{figure}[ht]
\begin{center}
\includegraphics[width=0.95\linewidth]
{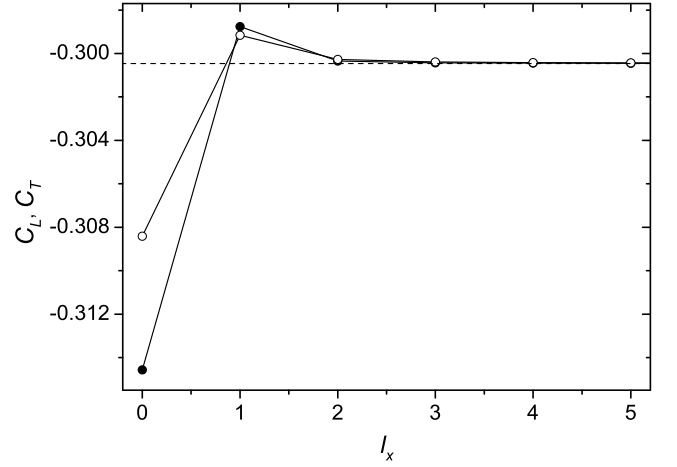} \caption{The nearest-neighbor spin correlations parallel (filled circles) and perpendicular (open circles) to the surface as functions of the distance $l_x$ from the surface for $T=0$. Solid lines are the guide to the eye. The dashed line indicates the bulk value.} \label{fig2}
\end{center}
\end{figure}

The calculated values of these spin correlations are shown in Fig.~\ref{fig2}. As seen from the figure, main deviations from the bulk value of the correlations fall on the surface and the second to the surface layers, i.e. on the existence domain of the surface mode. The largest in absolute value spin correlations are observed on the surface and between the surface and the second to the surface layers. This observation conforms with the quasi-two-dimensional character of the surface mode and the fact that the modulus of the nearest-neighbor spin correlation in the infinite 2D antiferromagnet ($|\!\left\langle\e{S_l} \e{S_{l+a}}\right\rangle\!|=0.329$ \cite{metlitski}) exceeds its value in the 3D case ($|\!\left\langle\e S_{\e l l_x}\e S_{\e l+\e a,l_x}\right\rangle\!|=0.3005$ in our calculations, which is close to the values obtained earlier \cite{oitmaa}). Thus, we relate the increased spin correlations in the mentioned region to the surface spin-wave mode. The correlations in the second and between the second and the third layers are smaller in absolute value than the bulk value. This is connected with the destructive contributions of the surface and bulk modes. Qualitatively the obtained picture of spin correlations is similar to that observed in the 2D case \cite{hoglund,metlitski}. In the 3D case the deviations from the bulk value are smaller than in the 2D case -- the largest deviation is less than 5{\%} in comparison with 12{\%} in the latter case. Besides, in the 3D case the influence of the edge decays more rapidly with distance from it -- already in the third layer from the surface the correlations differ only slightly from the bulk value. Again we notice that in this and deeper layers surface excitations are hardly perceptible in Green's function and the spectrum is dominated by bulk modes.
\begin{figure}[ht]
\begin{center}
\includegraphics[width=0.5\linewidth]
{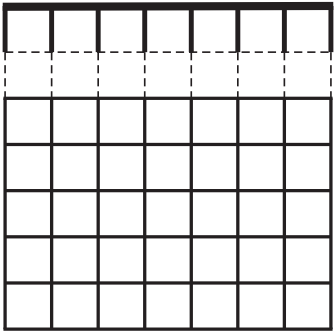} \caption{The pattern of nearest-neighbor spin correlations near the surface
of the 3D antiferromagnet. Thin solid lines indicate spin correlations
approximately equal to the bulk value, thick solid lines are larger in
absolute value correlations, while dotted lines correspond to correlations
smaller than the bulk value.}\label{fig3}
\end{center}
\end{figure}

The increased correlations near the surface lead to the comb structure shown in Fig.~\ref{fig3}. This structure is qualitatively similar to that obtained in the 2D case. From the above discussion it follows that the origin of such a spin correlation pattern is related to the existence of the surface spin-wave mode.

\section{Conclusion}
In this paper, we have studied the effect of the surface on the
nearest-neighbor spin correlations in the 3D spin-$\frac{1}{2}$
Heisenberg antiferromagnet on a simple cubic lattice. We have found that the spin correlations are enhanced on and close to the surface. We relate this enhancement to the existence of the surface spin-wave mode which ejects bulk modes from the region near the surface. The pattern of spin correlation in the 3D case is similar to that obtained recently in the 2D Heisenberg antiferromagnet. The parallels both in the mathematical description and in the physical behavior have been revealed between the considered problem and the problem of a local defect in the crystal.

\section*{Acknowledgements}
This work was supported by the ETF grant No. 6918.


\begin{thebibliography}{00}
\bibitem{mahajan} A.V.~Mahajan, H.~Alloul, G.~Collin, J.F.~Marucco, Phys.\ Rev.\ Lett.\ 72 (1994) 3100.
\bibitem{takigawa} M.~Takigawa, N.~Motoyama, H.~Eisaki, S.~Uchida, Phys.\ Rev.\ B 55 (1997) 14129.
\bibitem{hoglund} K.H.~H\"{o}glund, A.W.~Sandvik, Phys.\ Rev.\ B 79 (2009) 202405.
\bibitem{metlitski} M.A.~Metlitski, S.~Sachdev, Phys.\ Rev.\ B 78 (2008) 174410.
\bibitem{pardini} T.~Pardini, R.R.P.~Singh, Phys.\ Rev.\ B 79 (2009) 094413.
\bibitem{sherman} A.~Sherman, N.~Voropajeva, preprint arXiv:0904.4314.
\bibitem{lifshits} I.M.~Lifshits, Soviet Phys.\ Uspekhi 7 (1965) 549.
\bibitem{tyablikov} S.V.~Tyablikov, Methods of the Quantum Theory of Magnetism, Plenum Press, New York, 1967.
\bibitem{oitmaa} J.~Oitmaa, C.J.~Hamer, Z.~Weihong, Phys.\ Rev.\ B 50 (1994) 3877.
\end{thebibliography}
\end{document}